\documentclass[preprint,amsmath,amssymb,11pt,english,aps,showpacs,showkeys]{revtex4}
\usepackage{graphicx}
\usepackage{graphics}
\usepackage{amsmath}
\usepackage{dcolumn}
\usepackage{amssymb}
\usepackage{bm}
\usepackage[latin1]{inputenc}

\begin{document}
\title{Relativistic Landau-Aharonov-Casher quantization based on the Lorentz symmetry violation background}
\author{K. Bakke}
\email{kbakke@fisica.ufpb.br}
\affiliation{Departamento de F\'isica, Universidade Federal da Para\'iba, Caixa Postal 5008, 58051-970, Jo\~ao Pessoa, PB, Brazil.}

\author{H. Belich} 
\affiliation{Universidade Federal do Esp\'{\i}rito Santo, Departamento de F\'{\i}sica e Qu\'{\i}mica, Av. Fernando Ferrari, 514, Goiabeiras, 29060-900, Vit\'{o}ria, ES, Brazil.}

\author{E. O. Silva}
\affiliation{Departamento de F\'{\i}sica, Universidade Federal Maranh\~{a}o, Campus Universit\'{a}rio do Bacanga, 65085-580, S\~{a}o Lu\'{\i}s, MA, Brazil.}

\begin{abstract}
Based on the discussions about the Aharonov-Casher effect in the Lorentz symmetry violation background, we show that the analogue of the relativistic Landau quantization in the Aharonov-Casher setup can be achieved in the Lorentz-symmetry violation background.

\end{abstract}


\maketitle

\section{Introduction}

The quantum dynamics of permanent dipole moments has been widely studied in recent years. Anandan \cite{anan} and Silenko \cite{sil} made interesting discussions about the quantum effects of the interaction of the permanent dipole moments. Aharonov and Casher \cite{ac} showed that the interaction between the permanent magnetic moment of a neutral particle with a radial electric field provides the appearance of a geometric quantum phase on the wave function of the neutral particle. The appearance of this geometric quantum phase has been denoted the Aharonov-Casher effect \cite{ac}. Recently, the Aharonov-Casher effect has been widely discussed in the literature, for instance, in relation to the topological nature \cite{ac4}, the nonlocality and the nondispersivity \cite{ac6,disp,disp2}. Analogous effects to the Aharonov-Casher effect have been studied in the noncommutativity quantum mechanics \cite{fur1}, in the presence of topological defects \cite{bf1}, in noninertial reference frames \cite{bf3}, in holonomic quantum computation \cite{es}, and in the Lorentz symmetry violation background \cite{lin4,belich}. He, McKellar and Wilkens \cite{hmw} proposed the dual effect of the Aharonov-Casher effect by considering a neutral particle with a permanent electric dipole moment interacting with a magnetic field. This dual effect is known as the He-McKellar-Wilkens effect \cite{hmw}. The solid state analogue of the He-McKellar-Wilkens effect has been studied in \cite{fur2} and more studies of quantum phases for electric and magnetic dipoles have been made in \cite{fur3}.

In this paper, based on the discussions about the Aharonov-Casher effect in the Lorentz-symmetry violation background \cite{lin4,belich}, we wish to study the relativistic Landau-Aharonov-Casher quantization \cite{bf5} in the Lorentz-symmetry violation background. The relativistic Landau-Aharonov-Casher quantization has been proposed in \cite{bf5} and it has been also studied in a topological defect spacetime \cite{bf6}. The relativistic Landau-Aharonov-Casher quantization has been based on the field configuration conditions of the nonrelativistic Landau quantization in the Aharonov-Casher setup proposed in Ref. \cite{er}. The nonrelativistic Landau quantization in the Aharonov-Casher setup has also been studied in the presence of topological defects \cite{bf4} and in a noninertial frame \cite{bf16}. The Landau quantization has been extended to the He-McKellar-Wilkens setup \cite{lin}, the noncommutative space \cite{lin2}, a noninertial frame \cite{b}, and the Lorentz symmetry violation background and supersymmetric quantum mechanics \cite{lin5}. Another study of the Landau quantization for neutral particle has been made for an induced electric dipole \cite{lin3}. In this work, we discuss the analogue of the relativistic Landau-Aharonov-Casher quantization in the Lorentz symmetry violation background. We consider a Lorentz symmetry violation background based on a space-like vector, where we can define the magnetic dipole moment of the neutral particle as $\vec{\mu}=g\,\vec{b}$ \cite{belich}. In this way, we show that the relativistic Landau quantization can be achieved.

The structure of this paper is: in section II, we present the mathematical tools and discuss the relativistic Landau-Aharonov-Casher quantization in the Lorentz symmetry violation background; in section III, we present our conclusions.

\section{the Analogue of the relativistic Landau-Aharonov-Casher quantization}

We start this section by introducing the Lorentz symmetry violation background and the mathematical tools that we need to work in curvilinear coordinates. In the following, we consider a Lorentz symmetry violation background based on a space-like vector and show the relativistic analogue of the relativistic Landau-Aharonov-Casher quantization \cite{bf5}. At the end, we discuss the nonrelativistic limit of the energy levels and obtain the Dirac spinors for positive-energy solutions. 

The Lorentz symmetry violation background is described by the introduction of a nonminimal coupling in the Dirac equation given by $i\gamma^{\mu}\,\partial_{\mu}\rightarrow\,i\gamma^{\mu}\,\partial_{\mu}-g\,b^{\nu}\,\tilde{F}_{\mu\nu}\,\gamma^{\mu}$ \cite{carroll}, where the 4-vector $b^{\nu}$ is a fixed vector and acts on a vector field which breaks the Lorentz symmetry and $\tilde{F}_{\mu\nu}=\frac{1}{2}\epsilon_{\mu\nu\alpha\beta}\,F^{\alpha\beta}$ is the dual electromagnetic tensor. The tensor $F^{\mu\nu}$ is the electromagnetic field tensor $F^{0i}=-F^{i0}=E^{i}$, $F^{ij}=-F^{ji}=\epsilon^{ijk}\,B_{k}$ \cite{carroll2}. This proposal \cite{carroll} suggests that this background field, intervening in spacetime, may correct or generate some new properties to the particles. Then, independently of
the spin of the particle, the Lorentz-symmetry violating background vector may yield contribution to
the magnetic moment and the Aharonov-Casher phase of the particle, even if it is electrically neutral; whenever a particular nonminimal coupling of the particle to the electromagnetic field and the Lorentz-breaking vector is considered \cite{belich}. Everything goes as if the Lorentz-symmetry violating background endows each elementary particle, even those spinless and electrically neutral, with a universal contribution to its magnetic moment and, consequently, to its Aharonov-Casher phase \cite{belich}.

In order to discuss the relativistic Landau-Aharonov-Casher quantization in the Lorentz symmetry violation background, we work with the units $\hbar=c=1$ in the flat spacetime background, but in curvilinear coordinates because the Aharonov-Casher setup is based on the cylindrical symmetry. In this way, we can write the line element of the Minkowski spacetime background in cylindrical coordinates: $ds^{2}=-dt^{2}+d\rho^{2}+\rho^{2}d\varphi^{2}+dz^{2}$. Thus, to treat this relativistic problem we should write the Dirac equation in cylindrical coordinates. It has been shown in Ref. \cite{schu} that, after we apply a coordinate transformation $\frac{\partial}{\partial x^{\mu}}=\frac{\partial \bar{x}^{\nu}}{\partial x^{\mu}}\,\frac{\partial}{\partial\bar{x}^{\nu}}$ and a unitary transformation on the wave function $\psi\left(x\right)=U\,\widetilde{\psi}\left(\bar{x}\right)$, the Dirac equation can be written in any orthogonal system in the form:
\begin{eqnarray}
i\,\gamma^{\mu}\,D_{\mu}\,\psi+\frac{i}{2}\,\sum_{k}\,\gamma^{k}\,\left[D_{k}\,\ln\left(\frac{h_{1}\,h_{2}\,h_{3}}{h_{k}}\right)\right]\psi-m\psi=0,
\label{1.0}
\end{eqnarray}
where $D_{\mu}=\frac{1}{h_{\mu}}\,\partial_{\mu}$ is the derivative of the corresponding coordinate system, and $h_{k}$ corresponds to the scale factors of this coordinate system. For instance, in cylindrical coordinates, the scale factors are $h_{1}=1$, $h_{2}=\rho$, and $h_{3}=1$. Another way to treat this spin system in curvilinear coordinates is using the approach of the quantum field theory in curved spacetime \cite{bf4,bd,carroll2,bf5}. In a curved spacetime background, the rules of coordinate transformations for vectors, tensors and spinors must obey the rules established in general relativity. It is well-known in general relativity \cite{weinberg,bd} that, under a general coordinate transformation, the spinor representation of the Lorentz group either can exist or cannot exist. To incorporate the spinors into the general relativity, one should use the Principle of Equivalence to define locally inertial frames where the spinor representation of the Lorentz group is given as in the Minkowski spacetime. In this way, the spinors are defined locally, that is, the spinors are defined in the local reference frame of the observers, and transform according the rule: $\psi'\left(x\right)=D\left(\Lambda\left(x\right)\right)\,\psi\left(x\right)$, where $D\left(\Lambda\left(x\right)\right)$ is the spinor representation of the infinitesimal Lorentz group, and $\Lambda\left(x\right)$ corresponds to the local Lorentz transformations. But, how we can build a local reference frame for observers? A local reference frame for a observer can be build through a noncoordinate basis $\hat{\theta}^{a}=e^{a}_{\,\,\,\mu}\left(x\right)\,dx^{\mu}$, where the components $e^{a}_{\,\,\,\mu}\left(x\right)$ satisfy the relation $g_{\mu\nu}\left(x\right)=e^{a}_{\,\,\,\mu}\left(x\right)\,e^{b}_{\,\,\,\nu}\left(x\right)\,\eta_{ab}$ \cite{bd,naka}. The tensor $\eta_{ab}=\mathrm{diag}(- + + +)$ is the Minkowski tensor and the latin indices $a,b,c=0,1,2,3$ indicate the local reference frame of the observers. The components of the noncoordinate basis $e^{a}_{\,\,\,\mu}\left(x\right)$ are the well-known the \textit{tetrads} or \textit{Vierbein}. The inverse of the tetrads are defined as $dx^{\mu}=e^{\mu}_{\,\,\,a}\left(x\right)\,\hat{\theta}^{a}$, where they satisfy $e^{a}_{\,\,\,\mu}\left(x\right)\,e^{\mu}_{\,\,\,b}\left(x\right)=\delta^{a}_{\,\,\,b}$ and $e^{\mu}_{\,\,\,a}\left(x\right)\,e^{a}_{\,\,\,\nu}\left(x\right)=\delta^{\mu}_{\,\,\,\nu}$. There are many different ways to define the local reference of the observers in such a way the above relations are satisfied, for instance, we can choose $\hat{\theta}^{0}=dt;\,\,\,\hat{\theta}^{1}=d\rho;\,\,\,\hat{\theta}^{2}=\rho\,d\varphi;\,\,\,\hat{\theta}^{3}=dz$. But, since we are working with local frames, any information from the general coordinate system can be recovered through the derivatives \cite{weinberg}. An ordinary derivative transforms under a coordinate transformation $x\rightarrow x'$, that is, $\frac{\partial}{\partial x^{\mu}}=\frac{\partial \bar{x}^{\nu}}{\partial x^{\mu}}\,\frac{\partial}{\partial \bar{x}^{\nu}}$. Thus, locally, an ordinary derivative is defined as: $\frac{\partial}{\partial x^{a}}=e^{\mu}_{\,\,\,a}\left(x\right)\frac{\partial}{\partial x^{\mu}}$. In this way, to preserve the spinor representation of the infinitesimal Lorentz group, the derivative of a spinor in a general coordinate system must be given in the form:
\begin{eqnarray}
\partial_{a}\,\psi'\left(x\right)&=&\Lambda^{b}_{\,\,\,a}\left(x\right)\,e^{\mu}_{\,\,\,b}\left(x\right)\,\partial_{\mu}\left\{D\left(\Lambda\left(x\right)\right)\,\psi\left(x\right)\right\}\nonumber\\
[-3mm]\label{a}\\[-3mm]
&=&\Lambda^{b}_{\,\,\,a}\left(x\right)\,e^{\mu}_{\,\,\,b}\left(x\right)\,\left\{D\left(\Lambda\left(x\right)\right)\,\partial_{\mu}\,\psi\left(x\right)+\left[\partial_{\mu}D\left(\Lambda\left(x\right)\right)\right]\,\psi\left(x\right)\right\}.\nonumber
\end{eqnarray} 

The expression (\ref{a}) gives rise to the covariant derivative of a spinor, where the second term in (\ref{a}) is called the spinorial connection \cite{bd,weinberg}. Note that the second term of Eq. (\ref{1.0}) also corresponds to the spinorial connection. Hence, we have seen two alternative ways to write the Dirac equation in a general coordinate system, where the partial derivative is changed by the covariant derivative of a spinor. After some calculations, we can write the components of the covariant derivative of a spinor given in Eqs. (\ref{1.0}) and (\ref{a}) in the form: $\nabla_{\mu}=\partial_{\mu}+\Gamma_{\mu}\left(x\right)$, where $\Gamma_{\mu}\left(x\right)=\frac{i}{4}\,\omega_{\mu ab}\left(x\right)\,\Sigma^{ab}$ corresponds to the spinorial connection cited above \cite{bd,naka,weinberg}. The object $\omega_{\mu ab}\left(x\right)$ is the 1-form connection called the spin connection, whose components can be obtained by solving the Maurer-Cartan structure equations in the absence of the torsion field $d\hat{\theta}^{a}+\omega^{a}_{\,\,\,b}\wedge\hat{\theta}^{b}=0$ (the symbol $\wedge$ means the wedge product and the operator $d$ is the exterior derivative \cite{naka}), $\omega^{a}_{\,\,\,b}=\omega_{\mu\,\,\,\,b}^{\,\,\,a}\left(x\right)\,dx^{\mu}$ and $\Sigma^{ab}=\frac{i}{2}\left[\gamma^{a},\gamma^{b}\right]$. The $\gamma^{a}$ matrices are the Dirac matrices given in the Minkowski spacetime background \cite{bjd,bd}, \textit{i.e.},
\begin{eqnarray}
\gamma^{0}=\hat{\beta}=\left(
\begin{array}{cc}
1 & 0 \\
0 & -1 \\
\end{array}\right);\,\,\,\,\,\,
\gamma^{i}=\hat{\beta}\,\hat{\alpha}^{i}=\left(
\begin{array}{cc}
 0 & \sigma^{i} \\
-\sigma^{i} & 0 \\
\end{array}\right);\,\,\,\,\,\,\gamma^{5}=\gamma_{5}=\left(
\begin{array}{cc}
0 & I \\
I & 0 \\	
\end{array}\right);\,\,\,\,\Sigma^{i}=\left(
\begin{array}{cc}
\sigma^{i} & 0 \\
0 & \sigma^{i} \\	
\end{array}\right),
\label{2.3}
\end{eqnarray}
with $I$ being the $2\times2$ identity matrix and $\vec{\Sigma}$ being the spin vector. The matrices $\sigma^{i}$ are the Pauli matrices and satisfy the relation $\left(\sigma^{i}\,\sigma^{j}+\sigma^{j}\,\sigma^{i}\right)=2\,\eta^{ij}$ ($i,j=1,2,3$). Thus, solving the Maurer-Cartan structure equations in the absence of a torsion field, we obtain $\omega_{\varphi\,\,\,2}^{\,\,\,1}\left(x\right)=-\omega_{\varphi\,\,\,1}^{\,\,\,2}\left(x\right)=-1$, and one non-null component of the spinorial connection $\Gamma_{\varphi}=-\frac{i}{2}\,\Sigma^{3}$ (It is easy to check that in Cartesian coordinates, all the components of the spinorial connection are null \cite{bf5,schu}). The $\gamma^{\mu}\left(x\right)$ matrices are related to the $\gamma^{a}$ matrices via $\gamma^{\mu}\left(x\right)=e^{\mu}_{\,\,\,a}\left(x\right)\gamma^{a}$, thus, we have that $i\gamma^{\mu}\,\Gamma_{\mu}=\frac{i}{2\rho}\,\gamma^{1}$ \cite{bf5}. Thus, the Lorentz symmetry violation background described by the introduction of a nonminimal coupling can be written by
\begin{eqnarray}
i\gamma^{\mu}\partial_{\mu}\rightarrow\,i\gamma^{\mu}\,\partial_{\mu}+i\,\gamma^{\mu}\,\Gamma_{\mu}\left(x\right)-g\,b^{\nu}\,\tilde{F}_{\mu\nu}\,\gamma^{\mu}.
\label{2.1}
\end{eqnarray}

In this way, the Dirac equation in the Lorentz symmetry violation background becomes
\begin{eqnarray}
i\frac{\partial\psi}{\partial t}=m\,\hat{\beta}\,\psi+\vec{\alpha}\cdot\left[\vec{\pi}-g\,b^{0}\,\vec{B}+g\left(\vec{b}\times\vec{E}\right)\right]\psi+g\,\vec{b}\cdot\vec{B}\,\psi,
\label{2.6}
\end{eqnarray}
where $\pi_{i}=-i\partial_{i}-i\Gamma_{i}$ and we have an effective potential vector $\vec{A}_{\mathrm{eff}}=b^{0}\,\vec{B}+\left(\vec{b}\times\vec{E}\right)$. We have two distinct situations to be analyzed: the first one is that when we consider a space-like vector $\vec{b}=\left(0,0,b^{3}\right)$. In this case, following the discussions of Ref. \cite{belich}, we can consider the magnetic dipole moment of the neutral particle as $\vec{\mu}=g\,\vec{b}$, and we have that the effective potential vector becomes $\vec{A}_{\mathrm{eff}}'=\left(\vec{b}\times\vec{E}\right)$ which is similar to the effective potential vector of the Aharonov-Casher setup discussed \cite{bf5}. Thus, the relativistic Landau-Aharonov-Casher quantization can be achieved in this case if we choose a field configuration that satisfies the electrostatic conditions, produces no torque $\vec{\tau}=\vec{\mu}\times\vec{B}$ on the dipole moment $\vec{\mu}=g\,\vec{b}$ of the neutral particle and the effective potential vector produces an uniform effective magnetic field $\vec{B}_{\mathrm{eff}}=\vec{\nabla}\times\vec{A}_{\mathrm{eff}}'$. The second case is given when we consider a time-like 4-vector $b^{\mu}=\left(b^{0},0,0,0\right)$. In this case, the effective potential vector becomes $\vec{A}_{\mathrm{eff}}''=b^{0}\,\vec{B}$. Since there is no analog of the magnetic dipole moment in this case, the relativistic Landau quantization can be achieved by taking into account the two of the conditions cited above: the field configuration must satisfy the electrostatic conditions and the existence of an uniform effective magnetic field given by $\vec{B}_{\mathrm{eff}}=\vec{\nabla}\times\vec{A}_{\mathrm{eff}}''$ in the relativistic dynamics of the neutral particle.

In this work, we intend to discuss the relativistic Landau-Aharonov-Casher quantization in the Lorentz symmetry violation background. In this way, we consider the space-like vector $\vec{b}=\left(0,0,b^{3}\right)$ and a field configuration given by the radial electric field $\vec{E}=\frac{\lambda\rho}{2}\,\hat{\rho}$, where $\lambda$ is an electric charge density and $\hat{\rho}$ is an unitary vector in the $\rho$ direction. With these choices, it is easy to check that the electrostatic conditions are satisfied and there is no torque on the dipole moment $\vec{\mu}=g\,\vec{b}$ of the neutral particle. Moreover, we have an uniform effective magnetic field given by $\vec{B}_{\mathrm{eff}}=\lambda\,b^{3}\,\hat{z}$. The Dirac equation (\ref{2.6}) becomes
\begin{eqnarray}
i\frac{\partial\psi}{\partial t}=m\,\hat{\beta}\,\psi-i\,\hat{\alpha}^{1}\left(\frac{\partial}{\partial\rho}+\frac{1}{2\rho}\right)\psi-i\frac{\hat{\alpha}^{2}}{\rho}\frac{\partial\psi}{\partial\varphi}-i\hat{\alpha}^{3}\frac{\partial\psi}{\partial z}+\frac{g\,b^{3}\,\lambda}{2}\,\rho\,\hat{\alpha}^{2}\,\psi.
\label{2.7}
\end{eqnarray}

The solution of the Dirac equation (\ref{2.7}) is given in the form $\psi=e^{-i\mathcal{E}t}\left(\phi\,\,\,\chi\right)^{T}$, where $\phi$ and $\chi$ are spinors of two components. Substituting $\psi$ into (\ref{2.7}), we obtain two coupled equations for $\phi$ and $\chi$, where the first coupled equation is
\begin{eqnarray}
\left(\mathcal{E}-m\right)\phi=\left[-i\sigma^{1}\frac{\partial}{\partial\rho}-\frac{i\sigma^{1}}{2\rho}-\frac{i\sigma^{2}}{\rho}\frac{\partial}{\partial\varphi}-i\sigma^{3}\frac{\partial}{\partial z}+\frac{g\,b^{3}\,\lambda}{2}\,\rho\,\sigma^{2}\right]\chi,
\label{2.8}
\end{eqnarray}
while the second coupled equation is 
\begin{eqnarray}
\left(\mathcal{E}+m\right)\chi=\left[-i\sigma^{1}\frac{\partial}{\partial\rho}-\frac{i\sigma^{1}}{2\rho}-\frac{i\sigma^{2}}{\rho}\frac{\partial}{\partial\varphi}-i\sigma^{3}\frac{\partial}{\partial z}+\frac{g\,b^{3}\,\lambda}{2}\,\rho\,\sigma^{2}\right]\phi.
\label{2.9}
\end{eqnarray}

Thus, eliminating $\chi$ in (\ref{2.9}) and substituting into (\ref{2.8}), we obtain the following second order differential equation
\begin{eqnarray}
\left(\mathcal{E}^{2}-m^{2}\right)\phi&=&-\frac{\partial^{2}\phi}{\partial\rho^{2}}-\frac{1}{\rho}\frac{\partial\phi}{\partial\rho}-\frac{1}{\rho^{2}}\frac{\partial^{2}\phi}{\partial\varphi^{2}}-\frac{\partial^{2}\phi}{\partial z^{2}}+\frac{i\sigma^{3}}{\rho^{2}}\frac{\partial\phi}{\partial\varphi}+\frac{1}{4\rho^{2}}\phi\nonumber\\
[-2mm]\label{2.10}\\[-2mm]
&-&igb^{3}\lambda\frac{\partial}{\partial\varphi}+\frac{gb^{3}\lambda}{2}\,\sigma^{3}+\frac{\left(gb^{3}\lambda\right)^{2}}{4}\,\rho^{2}.\nonumber
\end{eqnarray}

From (\ref{2.10}), we can see that $\phi$ is an eigenfunction of $\sigma^{3}$, whose eigenvalues are $s=\pm1$, that is, $\sigma^{3}\phi=\pm\phi=s\phi$. Moreover, we can easily check that right-hand-side of (\ref{2.10}) commutes with the operators $\hat{J}_{z}=-i\partial_{\varphi}+\sigma^{3}/2$ and $p_{z}=-i\partial_{z}$. Thus, we can write the solution of the second order equation (\ref{2.10}) in the form: $\phi_{s}=e^{i\left(l+\frac{1}{2}\right)\varphi}\,e^{ikz}\,R_{s}\left(\rho\right)$. Substituting this solution into (\ref{2.10}), we obtain the radial equation:
\begin{eqnarray}
\left[\frac{d^{2}}{d\rho^{2}}+\frac{1}{\rho}\frac{d}{d\rho}-\frac{\zeta_{s}^{2}}{\rho^{2}}-\frac{\left(gb^{3}\lambda\right)^{2}}{4}\,\rho^{2}+\beta_{s}\right]R_{s}=0,
\label{2.11}
\end{eqnarray}
where we have defined $\zeta_{s}=l+\frac{1}{2}\left(1-s\right)$ and $\beta_{s}=\mathcal{E}^{2}-m^{2}-gb^{3}\lambda\zeta_{s}-sgb^{3}\lambda$. Since there is no torque on the dipole moment, we have taken $k=0$ \cite{er} to define $\beta_{s}$. Let us make the coordinate transformation: $\xi=\frac{gb^{3}\lambda}{2}\,\rho^{2}$, thus, the radial equation (\ref{2.11}) becomes
\begin{eqnarray}
\left[\xi\,\frac{d^{2}}{d\xi^{2}}+\frac{d}{d\xi}-\frac{\zeta_{s}^{2}}{4\xi}-\frac{\xi}{4}+\frac{\beta_{s}}{2gb^{3}\lambda}\right]R_{s}=0.
\label{2.12}
\end{eqnarray}
The solution of (\ref{2.12}) is given by $R_{s}\left(\xi\right)=e^{-\xi/2}\,\xi^{\left|\zeta_{s}\right|/2}\,F_{s}$. Substituting this solution into (\ref{2.12}), we obtain
\begin{eqnarray}
\xi\frac{d^{2}F_{s}}{d\xi^{2}}+\left[\left|\zeta_{s}\right|+1-\xi\right]\frac{dF_{s}}{d\xi}+\left[\frac{\beta_{s}}{2gb^{3}\lambda}-\frac{\left|\zeta_{s}\right|}{2}-\frac{1}{2}\right]F_{s}=0.
\label{2.13}
\end{eqnarray}

The equation (\ref{2.13}) is the Kummer equation or the confluent hypergeometric equation \cite{abra}. In order to obtain a regular solution at the origin, we consider only the Kummer function of first kind given by $F_{s}\left(\xi\right)=F\left[\frac{\left|\zeta_{s}\right|}{2}+\frac{1}{2}-\frac{\beta_{s}}{2gb^{3}\lambda},\left|\zeta_{s}\right|+1,\xi\right]$ \cite{abra}. By imposing the condition where the confluent hypergeometric series becomes a polynomial of degree $n$, where $n=0,1,2,\ldots$, the radial wave function of the neutral particle becomes finite everywhere \cite{landau2} and it allow us to obtain the relativistic energy levels. This can be possible if $\frac{\left|\zeta_{s}\right|}{2}+\frac{1}{2}-\frac{\beta_{s}}{2gb^{3}\lambda}=-n$. In this way, with $\beta_{s}=\mathcal{E}^{2}-m^{2}-gb^{3}\lambda\zeta_{s}-sgb^{3}\lambda$, we have
\begin{eqnarray}
\mathcal{E}_{n,\,l}=\sqrt{m^{2}+2gb^{3}\lambda\left[n+\frac{\left|\zeta_{s}\right|}{2}+\frac{\zeta_{s}}{2}+\frac{1}{2}\left(1+s\right)\right]}.
\label{2.14}
\end{eqnarray}

The energy levels (\ref{2.14}) correspond to the analogue of the relativistic Landau-Aharonov-Casher quantization in the Lorentz symmetry violation background. This could be achieved by assuming that the magnetic dipole moment of the neutral particle is given by $\vec{\mu}=g\,\vec{b}$, where the vector $\vec{b}$ corresponds to a space-like vector of the Lorentz symmetry violation background which has been defined parallel to the $z$-axis of the spacetime. If we have had chosen another scenario for the Lorentz symmetry violation, the magnetic dipole moment of the neutral particle would have another direction breaking the Aharonov-Casher setup and making that the analogue of the relativistic Landau-Aharonov-Casher quantization could not be achieved anymore.

The nonrelativistic limit of the energy levels (\ref{2.14}) can be obtained by applying the Taylor expansion up to the first order term in (\ref{2.14}). Writing (\ref{2.14}) in the form  $\mathcal{E}_{n,\,l}=m\sqrt{1+\frac{2gb^{3}\lambda}{m^{2}}\left[n+\frac{\left|\zeta_{s}\right|}{2}+\frac{\zeta_{s}}{2}+\frac{1}{2}\left(1+s\right)\right]}$ and applying the Taylor expansion, we obtain
\begin{eqnarray}
\mathcal{E}_{n,\,l}\approx m+\frac{gb^{3}\lambda}{m}\left[n+\frac{\left|\zeta_{s}\right|}{2}+\frac{\zeta_{s}}{2}+\frac{1}{2}\left(1+s\right)\right],
\label{2.18}
\end{eqnarray}
where $m$ corresponds to the rest mass of the neutral particle and remaining terms of (\ref{2.18}) correspond to the nonrelativistic Landau levels for a neutral particle in the Lorentz symmetry violation background. The cyclotron frequency corresponds to $\omega=\frac{gb^{3}\lambda}{m}$, which is analogous to the cyclotron frequency $\omega_{\mathrm{AC}}=\frac{\mu\lambda}{m}$ since we have that the magnetic dipole moment of the neutral particle is given by $\vec{\mu}=g\,\vec{b}$ in the Lorentz symmetry violation background. This nonrelativistic Landau quantization for a neutral particle has been discussed in \cite{lin5} by applying the Foldy-Wouthuyssen transformation \cite{fw}. Hence, we can see that by applying the Taylor expansion (up to the first order term) to the expression of the relativistic energy levels (\ref{2.14}), we can obtain the nonrelativistic Landau-Aharonov-Casher quantization.

At this moment, we intend to obtain the Dirac spinors. In order to obtain the appropriate solutions of the Dirac equation (\ref{2.7}), we must solve the system of coupled equation given in (\ref{2.8}) and (\ref{2.9}). Until now, we obtained the solutions for the two-spinor $\phi$, where the radial eigenfunctions are $\phi_{s}\left(\rho\right)=e^{-\frac{gb^{3}\lambda}{2}\,\rho^{2}}\,\left(\frac{gb^{3}\lambda}{2}\,\rho^{2}\right)^{\frac{\left|\zeta_{s}\right|}{2}}\,F\left[-n,\left|\zeta_{s}\right|+1,\frac{gb^{3}\lambda}{2}\,\rho^{2}\right]$.
Substituting these solutions into the equation (\ref{2.9}), we can obtain the solutions for the two-spinor $\chi$. Hence, the positive-energy solutions of the Dirac equation (\ref{2.7}) becomes
\begin{eqnarray}
\psi_{+}&=&\frac{f_{+}}{\left[\mathcal{E}+m\right]}\,\left(
\begin{array}{c}
1 \\
0\\
0\\
i\left[gb^{3}\lambda\rho-\frac{\left|l\right|}{\rho}+\frac{l}{\rho}\right]\\	
\end{array}\right)\,F\left[-n,\left|l\right|+1,\frac{gb^{3}\lambda}{2}\,\rho^{2}\right]\nonumber\\
[-3mm]\label{2.15}\\[-3mm]
&+&\frac{f_{+}}{\left[\mathcal{E}+m\right]}\,\,\left(
\begin{array}{c}
0\\
0\\
0\\
\frac{i\,gb^{3}\lambda\,\rho\,n}{\left[\left|l\right|+1\right]}\\	
\end{array}\right)F\left[-n+1,\left|l\right|+2,\frac{gb^{3}\lambda}{2}\,\rho^{2}\right],\nonumber
\end{eqnarray}
which is the parallel component of the Dirac spinor to the $z$ axis of the spacetime. The antiparallel component of the Dirac spinor is
\begin{eqnarray}
\psi_{-}&=&\frac{f_{-}}{\left[\mathcal{E}+m\right]}\,\left(
\begin{array}{c}
0 \\
1\\
i\,\left[\frac{\left|l+1\right|}{\rho}-\frac{l+1}{\rho}\right]\\	
0\\
\end{array}\right)\,F\left[-n,\left|l+1\right|+1,\frac{gb^{3}\lambda}{2}\,\rho^{2}\right]\nonumber\\
[-3mm]\label{2.16}\\[-3mm]
&+&\frac{f_{-}}{\left[\mathcal{E}+m\right]}\,\left(
\begin{array}{c}
0\\
0\\
\frac{i\,gb^{3}\lambda\,\rho\,n}{\left[\left|l+1\right|+1\right]}\\	
0\\
\end{array}\right)\,F\left[-n+1,\left|l+1\right|+2,\frac{gb^{3}\lambda}{2}\,\rho^{2}\right],\nonumber
\end{eqnarray}
where we have defined $f_{\pm}$ in equations (\ref{2.15}) and (\ref{2.16}) as $f_{\pm}=C\,e^{-i\mathcal{E}t}\,e^{i\left(l+\frac{1}{2}\right)\varphi}\,e^{-\frac{gb^{3}\lambda}{2}\,\rho^{2}}\,\left(\frac{gb^{3}\lambda}{2}\,\rho^{2}\right)^{\frac{\left|\zeta_{\pm}\right|}{2}}$.
The spinors (\ref{2.15}) and (\ref{2.16}) correspond to the positive-energy solutions of the Dirac equation (\ref{2.7}). One can use the same procedure to obtain the negative solutions of the Dirac equation (\ref{2.7}).

\section{Conclusions}

We have shown that the analogue of the relativistic Landau quantization in the Aharonov-Casher setup can be achieved in the Lorentz symmetry violation background. We have obtained the analogue of the relativistic Landau-Aharonov-Casher quantization by assuming a Lorentz symmetry violation background given by a space-like vector parallel to the z-axis of the spacetime and by considering the magnetic dipole moment of the neutral particle as being $\vec{\mu}=g\,\vec{b}$. In this way, we have shown that the field configuration conditions given by the electrostatic conditions, the absence of torque on the dipole moment and the presence of a uniform effective magnetic field, can be applied in this Lorentz symmetry violation scenario and the relativistic Landau-Aharonov-Casher quantization can be obtained. Finally, we have calculated the Dirac spinors for positive-energy solutions and discussed the nonrelativistic limit of the energy levels. 

This work has been partially supported by the Brazilian agencies CNPq, CAPES-PNPD and FAPEMA. 


\end{document}